\documentclass[journal]{IEEEtran}
\usepackage{cite}
\usepackage{amsmath,amssymb,amsfonts}
\usepackage{algorithmic}
\usepackage{graphicx}
\usepackage{textcomp}
\usepackage{wrapfig,colortbl}
\usepackage{multirow}
\usepackage{subcaption}
\usepackage{gensymb}

\def\BibTeX{{\rm B\kern-.05em{\sc i\kern-.025em b}\kern-.08em
    T\kern-.1667em\lower.7ex\hbox{E}\kern-.125emX}}
    
\begin{document}
\title{A Wavefield Correlation Approach to Improve Sound Speed Estimation in Ultrasound Autofocusing}
\author{Louise Zhuang, Samuel Beuret, Ben Frey, Saachi Munot, Walter Simson, Dongwoon Hyun, and Jeremy J. Dahl
\thanks{This work was supported in part by
the National Science Foundation Graduate Research Fellowship under Grant No. DGE-1656518 and National Institutes of Health Grant R01-027100.}
\thanks{Louise Zhuang is with the Department of Electrical Engineering, Stanford University, Stanford, CA 94305.}
\thanks{Samuel Beuret and Jeremy J. Dahl are with the Department of Radiology, School of Medicine, Stanford University, Stanford, CA 94305.}
\thanks{Ben Frey is with the Department of Applied Physics, Stanford University, Stanford, CA 94305.}
\thanks{Saachi Munot is with the Department of Biomedical Engineering, Columbia University, New York, NY 10027}
\thanks{Dongwoon Hyun and Walter Simson were with the Department of Radiology, School of Medicine, Stanford University at the initiation of this project and are now independently affliated.}
}

\maketitle

\begin{abstract}
    In pulse-echo ultrasound, aberration often degrades image quality when beamforming does not account for wavefront distortions. To address this issue, local sound speed estimators have been developed in the past decade for distributed aberration correction. Recently, methods based on iterative optimization have improved sound speed accuracy with respect to earlier approaches. However, the accuracy of these newer methods is limited by media with reverberation clutter and by the straight-ray model of wave propagation. To address these challenges, we propose using wavefield correlation (WFC) beamforming when performing sound speed optimization. WFC, an ultrasound adaptation of reverse time migration, correlates simulated forward-propagated transmit wavefields and backwards-propagated receive wavefields in order to reconstruct images. This process more accurately models wave propagation in heterogeneous media and can decrease diffuse clutter due to its spatiotemporal matched filtering effect. We implement herein a WFC beamformer using an auto-differentiation software and estimate the sound speed map by optimizing a regularized common-midpoint phase focusing criterion using gradient descent. This approach is compared to a previous method relying on delay and sum (DAS) with straight-ray time delay calculations on a variety of simulated, phantom, and \textit{in vivo} data with large sound speed variations and clutter. Results show that using WFC decreases sound speed estimation error, leading to improvements in resolution and contrast in the corrected image. In particular, these promising results have potential to improve pulse-echo imaging for challenging clinical scenarios. 
\end{abstract}


\section{Introduction}
In ultrasound B-mode imaging, aberrations occur when a beam is not optimally focused during beamforming, typically due to spatial variations in sound speed within the medium that differ from the constant sound speed assumed for image reconstruction. This impaired focus can cause degradation in the image quality, including decreased resolution, visibility, and contrast \cite{Ali2023Aberration}. Clinically, aberration can impact medical diagnosis, especially when imaging at high depths, such as in abdominal screening \cite{Deng2017Quantifying, Smereczynski2018Pitfalls} and with overweight or obese individuals. To alleviate these problems, aberration correction methods that compensate for the focusing errors caused by sound speed heterogeneity have been of high interest for many decades.

Recent state-of-the-art approaches for distributed aberration correction involve the so-called "sound-speed-aware" beamformer, where a local sound speed estimate of the imaged medium is obtained, followed by beamforming with time delays calculated using the sound speed estimate to better focus the image throughout the medium \cite{Jaeger2015Full}. Early pulse-echo local sound speed estimators that could be used for typical medical ultrasound applications were based on solving tomographic inverse problems \cite{Jaeger2015Computed, Sanabria2018Spatial, Stahli2020Improved, Rau2021Speed} or estimating the depthwise global (cumulative) sound speed and performing Dix inversion to compute the local sound speed per depth \cite{Jakovljevic2018Local, Ali2022Local}. These early methods produced novel sound speed estimation results for the traditionally challenging pulse-echo imaging configuration and greatly advanced knowledge of local sound speed estimation in pulse-echo ultrasound, but the estimates often were limited in resolution and accuracy, especially \textit{in vivo}. 
Nevertheless, from these estimates of the local sound speed, image reconstruction can be performed to improve the image focus, often through delay-and-sum (DAS) beamforming but with assumed straight-ray or bent-ray (refracted) propagation paths to compute time delays \cite{Wang2013Transcranial,Ali2018Distributed, Beuret2020Refraction}. Although these methods could be memory- or time-consuming when computing propagation paths at high resolutions, they often lead to significant improvements in visual quality, even with inaccuracies in the computed local sound speed maps \cite{Ali2022Distributed, Ali2023Iterative}. 

To improve the determination of local sound speed, iterative closed-loop versions of sound speed estimators have been proposed. An iterative version of pulse-echo tomographic sound speed estimation (iterative CUTE) was developed in \cite{Zengqiu2024Iterative}. The final predictions from this approach have increased accuracy compared to the prior CUTE model \cite{Stahli2020Improved} because repeatedly updating the phase shifts for the inversion process using the previously estimated sound speed map alleviates the estimation bias from the initial assumed sound speed. Likewise, \cite{Ali2023Iterative} improves upon \cite{Ali2022Distributed}, in part by iteratively calculating aberration delays from beamformed data and updating the sound speed estimate in a closed loop. 
While many of these estimators utilize the relationship between measured phase shifts, or aberration delays, caused by local sound speed variations in the medium to solve an inverse problem, some of the more recent aberration correction approaches optimize an image focusing criterion with respect to the local sound speed instead. In earlier decades, optimizing a focusing criterion was used for simpler phase-screen models of aberration \cite{Krishnan1996Adaptive, Masoy2005Iteration}, but this principle has recently been expanded to target distributed aberration correction as well \cite{Wang2025Pixel}. However, these approaches do not incorporate local sound speed to the image reconstruction process. 
Recently, we introduced ultrasound autofocusing, a distributed aberration correction technique that ties image focusing criteria to gradient-descent-based sound speed estimation to iteratively update a local sound speed estimate and apply DAS beamforming via delays calculated by a straight-ray model of wave propagation \cite{Simson2025Ultrasound}. In particular, this method has been shown to produce excellent distributed aberration correction. 

While sound speed estimation improved significantly with respect to prior models, the accuracy of ultrasound autofocusing for local sound speed imaging is still limited by several factors. First, estimation accuracy can degrade with data corrupted by acoustic noise. As shown in \cite{Jakovljevic2018Local}, diffuse reverberation can significantly increase the bias in sound speed estimation due to multiple scattering and increased clutter that decreases signal coherence. Additionally, many sound speed estimators utilize a straight- or bent-ray DAS beamforming model to calculate time delays which do not factor in diffraction effects, leading to inaccuracies in complex heterogeneous media or media with large sound speed changes.
However, an alternative beamforming strategy, referred hereafter as wavefield correlation (WFC), can be used to model more complex wave propagation such as diffraction. Wavefield correlation, also known as shot-profile and reverse-time migration in geophysics \cite{Schliecher2008shot}, consists of simulating the forward propagation of the transmit wavefields and backpropagation of the receive wavefields, followed by the correlation of those wavefields 
to form the final image 
\cite{Schwab2018Full, Ali2021Fourier, Ali2022Distributed}. This approach can more accurately represent wave propagation during the image reconstruction process and correct for aberration given an estimated local sound speed map \cite{Ali2022Distributed}. Due to the properties of the wavefield simulation and subsequent correlation, this method can also reduce the effect of diffuse reverberation in the final beamformed data \cite{Zhuang2024Simultaneous} by acting as a spatial matched filter. 

In this manuscript, we extend our prior work in \cite{Zhuang2024Simultaneous} to incorporate sound speed estimation based on image-quality optimization using WFC as the beamformer in the autofocusing approach \cite{Simson2025Ultrasound}, with the goal of improving local sound speed estimation in complex media. While this approach is similar to \cite{Ali2023Sound} in its usage of WFC as a beamformer and iteratively updating the sound speed model, our work differs fundamentally in the optimization criterion and how the sound speed estimate is updated. We review the background theory and the adaptation of WFC to the autofocusing model and then demonstrate the application of this method to simulation, phantom, and \textit{in vivo} imaging. 

\section{Theory}
\label{section:theory}
\begin{figure}[hb!]
    \centering
    \includegraphics[scale=0.40, trim={94mm, 26mm, 83mm, 22mm}, clip=true]{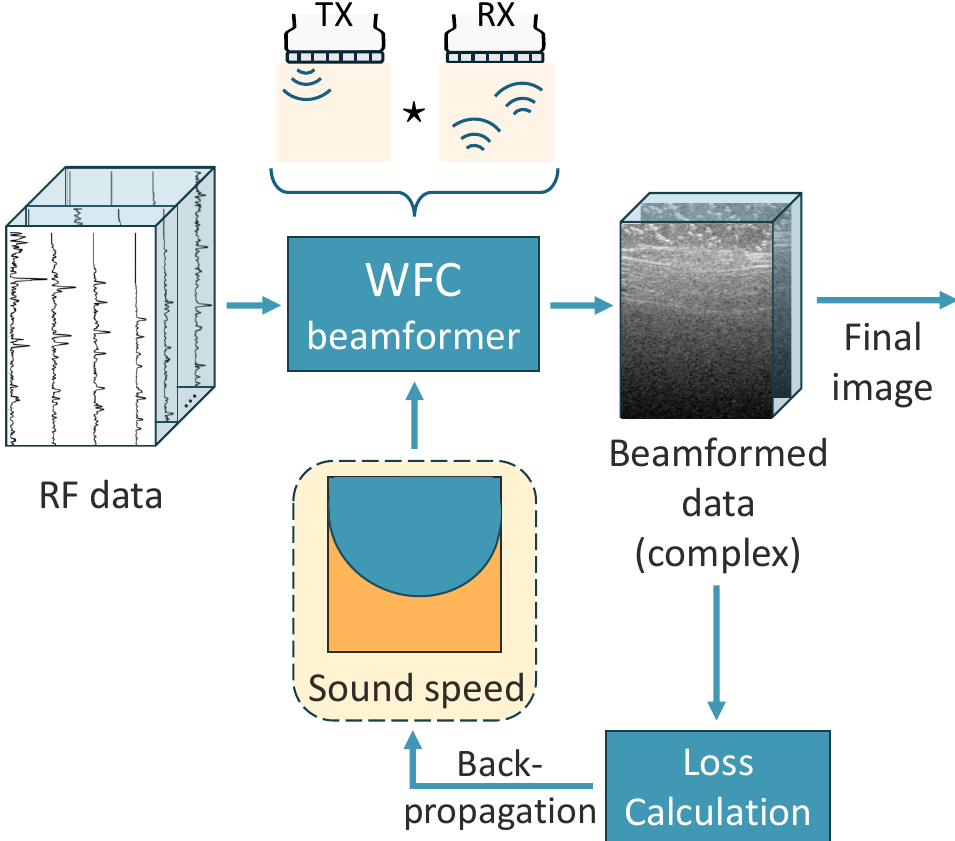}
    \caption{Diagram of the sound speed optimization approach. The common midpoint phase error is used as the image focusing criterion that we seek to optimize. The WFC beamforming process to obtain the focus criterion is tracked using auto-differentiation software, which is then used to backpropagate the loss gradient to update the local sound speed estimation. This iterative gradient-descent-based optimization process loops until a convergence criterion is met.}
    \label{fig:block_diagram}
\end{figure}

In this section, we explain the wavefield correlation beamforming approach, and describe how it is integrated into a sound speed optimization method to perform aberration correction and local sound speed estimation. The overall process used for estimating the local sound speed is illustrated in Figure \ref{fig:block_diagram}.

\subsection{Wavefield Correlation Beamformer}
Image reconstruction by correlating transmit and receive wavefields has previously been presented for ultrasound applications in \cite{Schwab2018Full, Rao2019Ultrasonic, Ali2021Fourier}. The implementation used in this work follows \cite{Ali2022Distributed}, where transmit pulses are forward propagated and the received data are backward propagated through heterogeneous media using a Fourier split-step simulator, and the two wavefields are correlated and summed to form each constant-depth row in the image. We briefly review this approach below, but we refer the reader to Schwab et al. \cite{Schwab2018Full}, Ali \cite{Ali2021Fourier}, and Ali et al. \cite{Ali2022Distributed} for a more detailed discussion.  

\subsubsection{Wavefield Simulations}
In the Wavefield Correlation (WFC) beamformer, the transmit ($p_{\text{tx}}(x, z, t)$) and receive ($p_{\text{rx}}(x, z, t)$) wavefields for azimuth dimension $x$, depth dimension $z$, and time dimension $t$ are required. These wavefields are obtained from the known or approximated transmit pulse $p_{\text{tx}}(x, z=0, t)$ and received channel data $p_{\text{rx}}(x, z=0, t)$ of the transducer, which are propagated via simulation. Transmit pulses are forward propagated through the heterogeneous media, while the received radiofrequency (RF) channel data are backward propagated through the same medium. The simulator used in this approach is a Fourier-domain split-step angular spectrum method (ASM), which is less computationally demanding than a full finite-difference-time-domain (FDTD) simulator typically used in reverse time migration. ASM can model wave propagation through a heterogeneous medium, albeit without considering multiple reflections \cite{Jones2016Comparison}. 
For simplicity, the implementation of ASM used herein does not incorporate attenuation or non-linearity. 

Let us restrict ourselves to linear transducers. To propagate a wavefield $p(x, z, t)$ using ASM, the wavefield is first transformed into the angular spectrum domain through a 2D Fourier transform:
\begin{equation}
    p(k_x, z, f) = \mathcal{F}_{x\longrightarrow k_x}\{\mathcal{F}_{t\longrightarrow f}[p(x, z, t)] \, .
\end{equation}
The wave is then propagated to the next depth sample by accounting for diffraction through a medium with a constant sound speed:
\begin{align}
    \begin{split}
    p_{\text{diffr.}}(k_x, z + \Delta z, f) =& \exp(\mp j2\pi \Delta z \sqrt{f^2 \bar{u}^2(z) - k_x^2}) \\ 
    &* p(k_x, z, f) \, ,
    \end{split}
\end{align}
where the $-$ sign is used for forward propagation, and the $+$ sign is used for backward propagation. Here, the slowness, $u(x, z) = 1 / c(x, z)$, for a medium with local sound speed $c(x,z)$ is decomposed as $u(x,z) = \bar{u}(z) + \Delta u(x,z)$ into a mean slowness and a differential slowness, respectively.
The mean slowness is used for this initial propagation step. 

Refraction during the propagation process from an inhomogeneous medium is then applied by taking the inverse Fourier transform in the azimuth dimension
\begin{equation}
    p_{\text{diffr.}}(x, z + \Delta z, f) = \mathcal{F}^{-1}_{k_x\longrightarrow x}[p_{\text{diffr.}}(k_x, z + \Delta z, f)]
\end{equation}
and multiplying the result by a phase term with the differential slowness
\begin{align}
    \begin{split}
    p(x, z + \Delta z, f) =& \exp(\mp j2\pi f \Delta z \Delta u(x,z)) \\
    &* p_{\text{diffr.}}(x, z + \Delta z, f) \, .
    \end{split}
\end{align}
These propagation steps are iterated through successive depths to obtain wavefields for the entire insonified medium.

\subsubsection{Image Reconstruction}
Wherever a scatterer is located in the medium, a wavefront peak will occur at that location in both the transmit and receive wavefields at the exact same time \cite{Claerbout1971Toward}. Thus, to reconstruct an image, $I(x,z)$, from these wavefields, the transmit and receive wavefields are correlated at each image location and summed over all frequencies
\begin{equation}
    I(x,z) = \int_{-\infty}^{\infty} p_{\text{tx}}(x, z, f) p_{\text{rx}}^*(x, z, f) df \, .
\end{equation}
This process is equivalent to integrating the correlations over all time instances \cite{Ali2021Fourier}. 

This beamforming approach has several advantages that benefits sound speed estimation and the resulting aberration correction. Through correlation of an ideal transmit wavefield with the backpropagated receive data, this beamforming process results in a spatio-temporal matched filtering effect, which can reduce diffuse reverberation noise in the resulting image \cite{Zhuang2024Simultaneous}. Additionally, since a local sound speed is inherently assumed during wavefield simulation, wave propagation with refractive and diffractive effects in heterogeneous media is more accurately modeled compared to other beamforming approaches, such as DAS using straight ray paths to calculate time delays. Finally, all steps taken during the beamforming process are differentiable, allowing 
integration into the autofocusing approach we have previously developed for simultaneous aberration correction and sound speed estimation \cite{Simson2025Ultrasound}. 

\subsection{Gradient-Descent-Based Sound Speed Estimation}
In this work, we rely on the gradient descent optimization approach proposed in \cite{Simson2025Ultrasound} for sound speed estimation. This method optimizes an image focusing metric relative to the local sound speed map, and updates the sound speed map in the same way as neural network loss backpropagation to minimize the loss function. All operations to calculate the final loss metric involving the local sound speed are tracked in a software package with automatic differentiation functionality, and the computed gradient value from these operations is used to update the local sound speed map. The process then repeats iteratively, much like neural network training, using the updated sound speed map to compute the loss metric and then auto-differentiation to obtain the gradient for updating the sound speed map.

The optimized loss metric used in this work is the sum of the common midpoint phase and a regularization term: 
\begin{equation}
    \mathcal{L}(c) = \mathcal{L}_{\text{CMPE}}(c) + \gamma \mathcal{R}_{\text{TV}}(c), 
    \label{eq:loss_fn}
\end{equation}
where $c$ represents the local sound speed map, $\mathcal{L}_{\text{CMPE}}$ represents the common midpoint phase error image focus metric. Here, we follow \cite{Simson2025Ultrasound} and use an anisotropic total variation regularizer $\mathcal{R}_{\text{TV}}$ with strength parameterized by a positive scalar regularization weight $\gamma$.

\subsubsection{Common Midpoint Phase Error}
The correlation of beamformed pressure signals at two receive locations $\boldsymbol{x_1}$, $\boldsymbol{x_2}$ is given by
\begin{equation}
    R(\boldsymbol{x_1}, \boldsymbol{x_2}) = \langle P(\boldsymbol{x_1}, \boldsymbol{x_2}) P^*(\boldsymbol{x_2})\rangle ,
\end{equation}
where $P$ denote the pressure signal and $\langle \cdot \rangle$ denotes cross-correlation.
Using the van Cittert-Zernike theorem, the correlation $R$ is formulated for pulse-echo ultrasound  through an aberrating layer according to \cite{Mallart1991van, Mallart1994Adaptive} as: 
\begin{align}
\label{eq:r_def}
    \begin{split}
        R(\boldsymbol{x_1}, \boldsymbol{x_2}) = & A \cdot e^{j[\phi(\boldsymbol{x_1}) - \phi (\boldsymbol{x_2})]} \cdot e^{j \frac{\pi}{\lambda z} (\boldsymbol{x_1 x_1} - \boldsymbol{x_2 x_2})} \\ & \cdot \int_U |P_i(\boldsymbol{u})|^2 e^{j2\pi  \frac{\boldsymbol{x_1} - \boldsymbol{x_2}}{\lambda z} \boldsymbol{u}} \mathop{} \!\mathrm{d}\boldsymbol{u}.
    \end{split}
\end{align}
In \eqref{eq:r_def}, $A$ is the resulting amplitude term, $\phi$ is the aberration expressed as a phase, $\lambda$ is the wavelength, $z$ is the depth, and $P_i$ is the transmitted pressure field, which insonifies the medium $U$. The integral term represents the Fourier transform of the source pressure field intensity, per the van Cittert-Zernike theorem. The second exponential term outside of the Fourier transform represents a location-dependent phase term based on the Fresnel approximation. The first exponential then represents any remaining phase difference between the two receive locations due to aberration. When the two receive locations share a common midpoint about the transducer axis, i.e., 
\begin{align}
    \boldsymbol{x_1} = \boldsymbol{x_m} + \Delta x ,\\
    \boldsymbol{x_2} = \boldsymbol{x_m} - \Delta x ,
\end{align}
and the data are beamformed, all phase terms except the first term cancel out or become deterministic. The first phase term thus accounts for any beamforming inaccuracies that cause aberration in the final beamformed data. Under perfect focusing, the first phase term should reduce to 0. The mean phase difference between common midpoint signals, called the common midpoint phase error \cite{Simson2025Ultrasound}, is proposed as a measure of the focusing quality of the beamformer and utilized as an optimization objective, i.e., 
\begin{equation}
    \mathcal{L}_{\text{CMPE}} = \angle R(\boldsymbol{x_1}, \boldsymbol{x_2}).
\end{equation}
In Simson et al. \cite{Simson2025Ultrasound}, the common midpoint phase error for the beamformed data over all common midpoint subapertures of size 17 elements was used after excluding values where the correlation coefficient at that location was below a threshold $\zeta$ 
to exclude noisy signals (e.g. from anechoic regions). Similarly, the present work uses this filtered version of the common midpoint phase error (CMPE), which is minimized during the optimization process. This metric is computed through software beamforming of the data using WFC. Separate simulations are performed for each subaperture considered, and both the ASM simulations and correlation process for final image formation are composed of differentiable operations. 

\subsubsection{Total Variation Regularization}
Because pulse-echo sound speed estimation is ill-posed, minimizing the CMPE by itself is not sufficient to obtain a physically meaningful local sound speed estimate. To regularize the objective function in the optimization problem, a weighted anisotropic total variation norm is used
\begin{equation}
    \mathcal{R}_{\text{TV}} = \sum_{i=2}^{M} \sum_{j=2}^{N} \alpha_x |D_x(i,j)|  + \alpha_z |D_z(i,j)|
    ,
    \label{eq:tvr}
\end{equation}
where
\begin{equation}
\begin{split}
    D_x(i, j) &= c(x_i,z_j) - c(x_{i-1}, z_j),\\
    D_z(i, j) &= c(x_i, z_j) - c(x_i, z_{j-1}),
\end{split}
\end{equation}
denote the forward first-order finite difference operators along the $x$ and $z$ axes applied to the discrete sound speed map $c$ of size $M \times N$. In addition, $\alpha_x$ and $\alpha_z$ denote in \eqref{eq:tvr} relative weights of the axial and lateral differences.

Total variation regularization is designed to favor piecewise-constant maps, with the rationale that the sound speed map obtained by minimizing Equation \eqref{eq:loss_fn} will be homogeneous within a specific tissue, thus removing potential distortions while still retaining sharp boundaries at the transitions between different tissues.

\section{Methods}
\label{section:methods}
\subsection{Simulations}
2D Fullwave \cite{Pinton2009Heterogeneous, Pinton2021Fullwave} simulations of a full synthetic aperture transmit sequence using a linear array were employed to evaluate the proposed approach with the transducer and transmission parameters included in Table \ref{tab:sim_trans}. Two simulations were performed that used numerical phantoms containing segmented abdominal wall media with acoustical properties described in \cite{Zhuang2025Labeled}, shrunken axially to reduce the total wall thickness to 10\,mm. The media below the wall then comprised of (a) a circular sound speed inclusion with a 5\,mm diameter and 1570\,m/s sound speed against a background of 1540\,m/s, and (b) a two layer medium with sound speeds of 1420\,m/s and 1555\,m/s. The density was set to 1000 kg/m$^3$ for the background region in simulation (a) and the first layer in simulation (b). The densities of the second layer and inclusion were calculated so the impedance would match the background and first layer, respectively, to avoid strong ``coherent'' reverberations. The abdominal wall tissue was assigned acoustical parameters based on \cite{Zhuang2025Labeled}. All other regions were simulated with a constant 0.10\,dB/MHz/cm attenuation and 0 nonlinearity. A Gaussian blur with radius 7.5 pixels was additionally applied to the acoustic parameter maps to reduce excessively large ``diffuse'' reverberations \cite{Dahl2014Reverb}. The simulations use 18 scatterers per resolution cell and use a grid spacing of 0.013\,mm per dimension. 

\begin{table}[ht!]
    \centering
    \caption{Simulated Transducer and Transmit Parameters}
    \label{tab:sim_trans}
    \begin{tabular}{c c}
        \textbf{Parameter} & \textbf{Value} \\ 
        \hline
        Transmit Center Frequency & 5.0\,MHz\\
        No. Cycles & 2 \\
        Fractional Bandwidth & 93.33\% \\
        No. Elements & 192 \\
        Element Pitch & 0.2\,mm \\
        Sampling Frequency & 32.08\,MHz
    \end{tabular}
\end{table}

\subsubsection{Phantom}
To evaluate the model's performance on phantom data, full synthetic aperture data were acquired from two types of phantoms using a Verasonics Vantage 256 scanner (Kirkland, WA) and a linear array. The first experiment re-used data from \cite{Ali2022Local}, where a chicken-breast layer between 10-15\,mm thick was placed above an ATS 549 phantom (Norfolk, VA). RF channel data were collected from a 192-element L12-3v array with a 5.95\,MHz transmit center frequency and a 25\,MHz sampling frequency. The ground truth sound speed in the chicken layer was measured using a piston transducer driven by a pulser-receiver, with time delays measured with an oscilloscope as described in the Appendix of \cite{Telichko2022Noninvasive}. This setup contained reverberation from the chicken-phantom boundary, as well as a sizable sound speed difference between the chicken (1568.5\,m/s) and the phantom (1450\,m/s). The data were demodulated and decimated by 2 prior to beamforming due to reduce memory usage. For the proposed WFC model, the data were then remodulated prior to the ASM simulation process.

In the second experiment, data from a fabricated gelatin phantom originally acquired for \cite{Ali2023Sound} were used to further analyze beamformer performance for heterogeneous media. Wire targets in the phantom serve as point targets, and cylindrical inclusions made of a gelatin-alcohol mixture created high sound speed regions relative to the background. An L12-5v, 50\,mm array was used to acquire the data with a 6.0\,MHz transmit center frequency and 25\,MHz sampling frequency. The sound speeds of the inclusions and background were not available for these data.

\subsubsection{\textit{In Vivo}}
\label{section:methods_invivo}
\textit{In vivo} data from \cite{Brickson2021Reverberation} and \cite{Telichko2022Noninvasive} were also used to evaluate the proposed method. 
From \cite{Brickson2021Reverberation}, thyroid and carotid artery data were obtained from three patients acquired using a 128-element full synthetic aperture sequence on an L12-3v transducer and Vantage 256 under a Stanford IRB approved protocol. A transmit center frequency of 5.95\,MHz was used, with a sampling frequency of 25\,MHz. Due to memory constraints, the data were demodulated and decimated by 2 prior to beamforming. These acquisitions contain reverberation clutter from superficial tissue that is visible in the vessel and anechoic nodules. 

Additional rat liver data from \cite{Telichko2022Noninvasive} was used to analyze sound speed estimation performance in tissues with more complex sound speed distributions. Data were originally captured with a full synthetic aperture sequence using 192 elements on an L12-3v transducer. The center transmit frequency was 7.81\,MHz, with a 31.25\,MHz sampling frequency. The global liver sound speeds were determined \textit{ex vivo} in a 37\degree C water tank using the same procedure as the meat-layer phantom data. In these data, speed of sound measurements from four of the liver lobes in a subset of these rats showed sound speed differences of up to 30\,m/s relative to the global liver measurements.

\subsection{Model Implementation}
\label{sxn:implementation}
The sound speed optimization pipeline with the proposed model is based on \cite{Simson2025Ultrasound} but factors in the WFC beamformer, as shown in Fig.~\ref{fig:block_diagram}, with CMPE threshold set to $\zeta=0.9$. We note that the proposed method is not a machine learning method, but employs tools such as automatic differentiation that were developed for machine learning applications. The proposed method was implemented using an Adam optimizer \cite{Kingma2017Adam} with default parameters and the loss function $\mathcal{L}$ in Equation \eqref{eq:loss_fn}. A learning rate scheduler was used to ramp up and decay the learning rate based on the iteration number. A linear warm-up was used starting from a learning rate of 0.005 and ending at a maximum learning rate of 0.015. After 110 iterations, the learning rate would then exponentially decay by a rate of 0.01275. 
For simulation and phantom experiments, the regularization parameter $\gamma$ was set to 0.3 to ensure sufficient suppression of artifacts in the sound speed map.
For the thyroid \textit{in vivo} acquisitions, where increased diffuse reverberation noise is present, the regularization weight was set to $\gamma = 0.75$ to more strongly enforce piecewise-constant sound speed estimates. For all data, the dimensional components of the total variation regularization in Equation \eqref{eq:tvr} were weighted with $\alpha_x = 5$ and $\alpha_z = 1$ to impose a more layer-wise sound speed structure, as in \cite{Simson2025Ultrasound}. 
The phase error was computed in 2D kernels, i.e., patches in which the phase is evaluated at certain finely spaced point locations, like phase calculations done in \cite{Stahli2020Improved, Simson2025Ultrasound}. The kernels were uniformly placed across the image field of view, with their centers arranged in a grid with 1\,mm spacing laterally and axially. Each individual kernel was of length $2\lambda$ in each dimension and had $2\lambda / 3$ intra-kernel point spacing laterally and axially. 
The sound speed maps were defined on a grid with a 1\,mm spacing along both axes. The axial and lateral ranges of the maps vary based on the experiment (see Section~\ref{section:results}) and differ from the field of view used in \cite{Simson2025Ultrasound} due to memory management with the WFC method. For display purposes only, the final sound speed estimates shown in this paper have been linearly interpolated to match the beamformed image grid. 
The initial constant sound speed used for beamforming and a constant time offset applied to alleviate global sound speed bias were simultaneously swept to optimize the CMPE metric before starting the first iteration. The sound speeds were swept from 1350 to 1750\,m/s with a step size of 2\,m/s, and the data start time offsets were swept from -4.9 to 0.3\,$\mu$s with a step size of 0.1\,$\mu$s. A finer phase error kernel grid with $2\lambda$ spacing in both dimensions was used for the initial sweep given the lack of memory constraints for the initialization. 
The optimization loop was run for 200 iterations for simulation data and 100 iterations for phantom and \textit{in vivo} data. These cutoffs were determined experimentally as the number of iterations for the sound speed maps to reach convergence. 

For the split step angular spectrum propagation used during WFC beamforming, the simulation grid has a spacing of one-half the element pitch laterally and $\lambda / 2$ axially, with a Butterworth anti-aliasing filter as in \cite{Ali2021Fourier} applied laterally during the propagation process to prevent wraparound artifacts. For consistency and memory constraints, the center 128 elements were used for beamforming all data, except for the rat acquisitions (see Section~\ref{section:methods_invivo}), which were beamformed using 160 elements to encompass a larger lateral span due to the shallower imaging depth. 

\subsubsection{Evaluation Criteria}
The optimization pipeline above was compared to the approach introduced in \cite{Simson2025Ultrasound}, which uses DAS with time delays calculated using a straight-ray travel path assumption through the medium. We otherwise kept the initialization values and optimizer setup described above for both the reference DAS and proposed WFC experiments. An f/\# of 1 was used to calculate the common midpoint phase error criterion when using DAS, and 48 points per ray were used for calculating the time delays. We compared the performance of ultrasound autofocusing with the WFC and DAS beamformers in the
simulated, phantom, and \textit{in vivo} data with reverberation noise and sound speed heterogeneities.

For simulation and phantom data with a known ground truth sound speed, the mean (bias), standard deviation of the error, and the mean absolute error (MAE) were computed. For data containing point targets or similar small, strong reflectors, the full width at half maximum (FWHM) is used to evaluate resolution. Additionally, for the \textit{in vivo} data with anechoic regions, the generalized contrast to noise ratio (gCNR) \cite{RodriguezMolares2020Generalized} is evaluated as a surrogate measure of image quality. 

\section{Results}
\label{section:results}
In our results, we refer to our models as the WFC and DAS models. However, in these comparisons, the reader should keep in mind that the DAS model should really be viewed as DAS with a straight-ray model of wave propagation and not simply the delay-and-sum process.

\subsection{Simulation}
\label{sec:res_sim}
\begin{figure*}[ht!]
    \centering
    \begin{subfigure}[b]{\textwidth}
        \centering
        \includegraphics[scale=0.36, trim={4mm, 3mm, 17mm, 5mm}, clip=true]{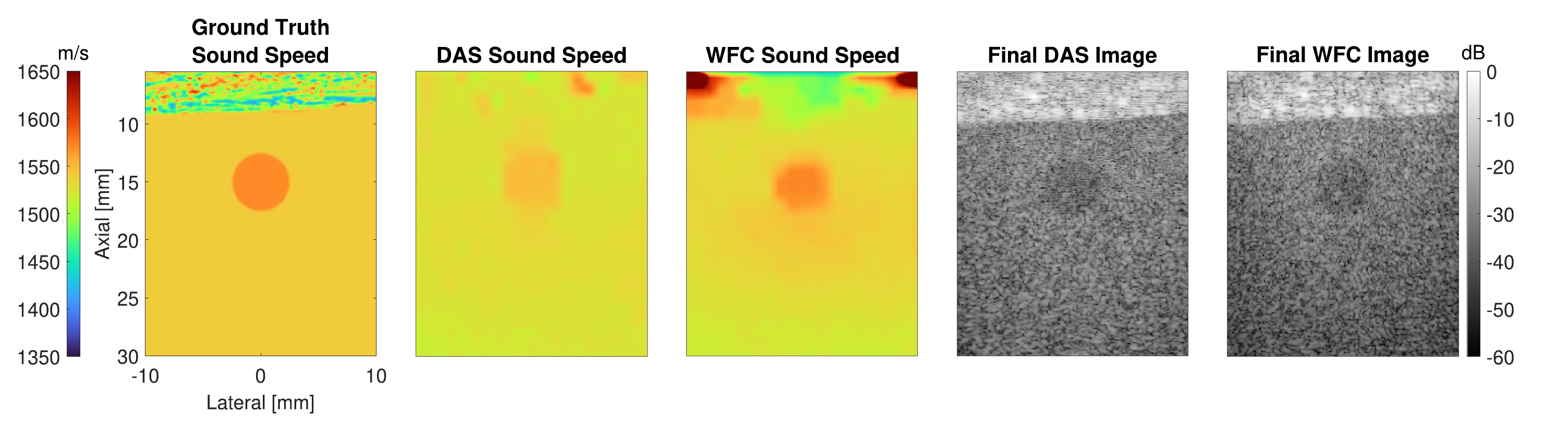}
        \caption{}
        \label{fig:abd_inclusion}
    \end{subfigure}
    \begin{subfigure}[b]{\textwidth}
        \centering
        \includegraphics[scale=0.36, trim={4mm, 3mm, 17mm, 5mm}, clip=true]{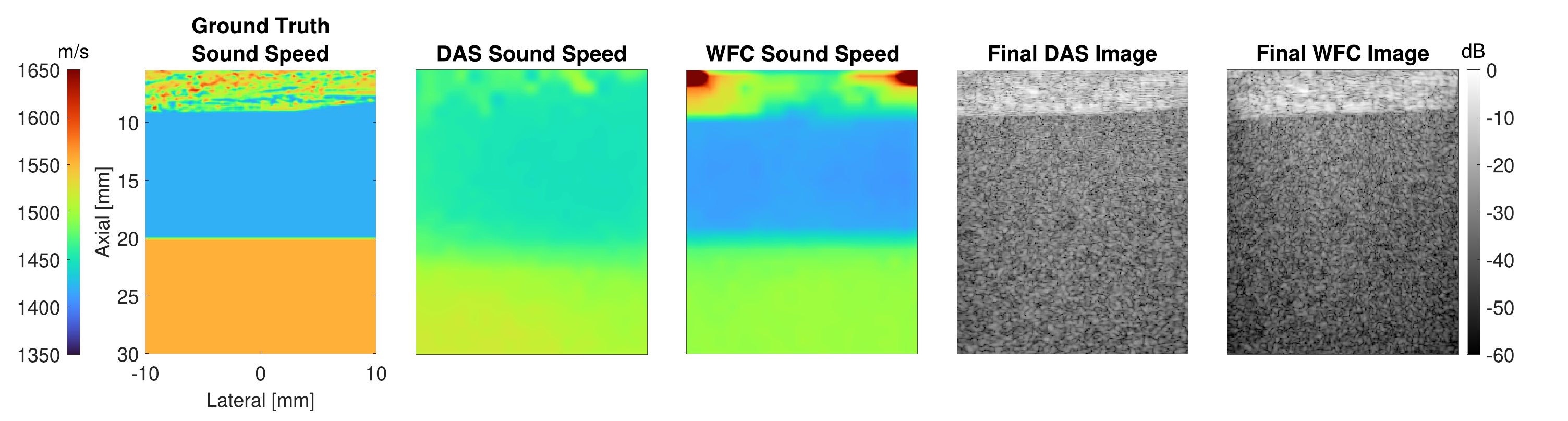}
        \caption{}
        \label{fig:abd_layer}
    \end{subfigure}
    \caption{Sound speed and final aberration-corrected image comparisons for abdominal wall simulations with a sound speed inclusion (a) and two sound speed layers (b), with the ground truth simulated sound speed shown on the leftmost panels. The sound speed estimates in the WFC model show clearer spatial differentiation of the abdominal wall, horizontal layers, and circular inclusion compared to the DAS estimates.}
    \label{fig:abd_sims}
\end{figure*}

The local sound speed estimates for the simulated inclusion (a) and layered media (b) under an abdominal wall are shown in Figure \ref{fig:abd_sims}, with sound speed errors for the regions underneath the abdominal wall quantified in Table \ref{tab:error_sim}. Most noticeably, the different sound speed regions are visually more distinguished with WFC compared to the straight-ray model with DAS. The reduction in bias with WFC is especially large for the circular sound speed inclusion in simulation (a) and the first, low sound speed layer in simulation (b). Although the bottom sound speed layer in (b) has a 10\,m/s higher bias for the WFC estimation, the sound speed variation is reduced compared to DAS, which has a higher error standard deviation. High sound speed corners, reminiscent of shallow-depth or corner artifacts sometimes seen in tomographic sound speed estimators \cite{Stahli2020Improved}, are visible at the top of the WFC estimates for both simulations. This corner artifact appears to focus energy inwards, away from the lateral edges of the WFC beamformed image. In addition, the shape of the circular inclusion in (a) is better recovered by the WFC model. The eccentricity of the inclusion---identified as the region with a sound speed higher than 6 standard deviation above the background sound speed---is reduced from 0.58 with the DAS model to 0.42 with the WFC model. Because a higher eccentricity represents a more oblong elliptical shape, the WFC inclusion is more circular and matches more closely to the expected inclusion shape.

\begin{table}[ht!]
    \centering
    \caption{Simulation Region Sound Speed Estimation Error}
    \label{tab:error_sim}
    \begin{tabular}{c|c|c|c|c}
        \textbf{Sim. Region} & \textbf{Beamformer} & \textbf{Bias} & \textbf{Std. Dev.} & \textbf{MAE} \\ 
        \hline
        \hline
        \multirow{2}{*}{(a) Inclusion} & DAS & -22.86 & 1.98 & 22.86 \\
         & WFC & -1.48 & 6.77 & 3.83 \\
         \hline
         \multirow{2}{*}{(a) Background} & DAS & -12.52 & 3.65 & 12.57 \\
         & WFC & -8.78 & 6.47 & 9.19 \\
         \hline 
         \hline
         \multirow{2}{*}{(b) Top Layer} & DAS & 32.22 & 4.66 & 32.22 \\
         & WFC & -3.74 & 6.92 & 5.92 \\
         \hline
         \multirow{2}{*}{(b) Bottom Layer} & DAS & -52.64 & 14.06 & 52.64 \\
         & WFC & -62.04 & 6.23 & 62.04 \\
    \end{tabular}
\end{table}

\subsection{Phantom}
\begin{figure*}[ht!]
    \centering
    \includegraphics[width=\linewidth, trim={4mm, 3mm, 17mm, 4mm}, clip=true]{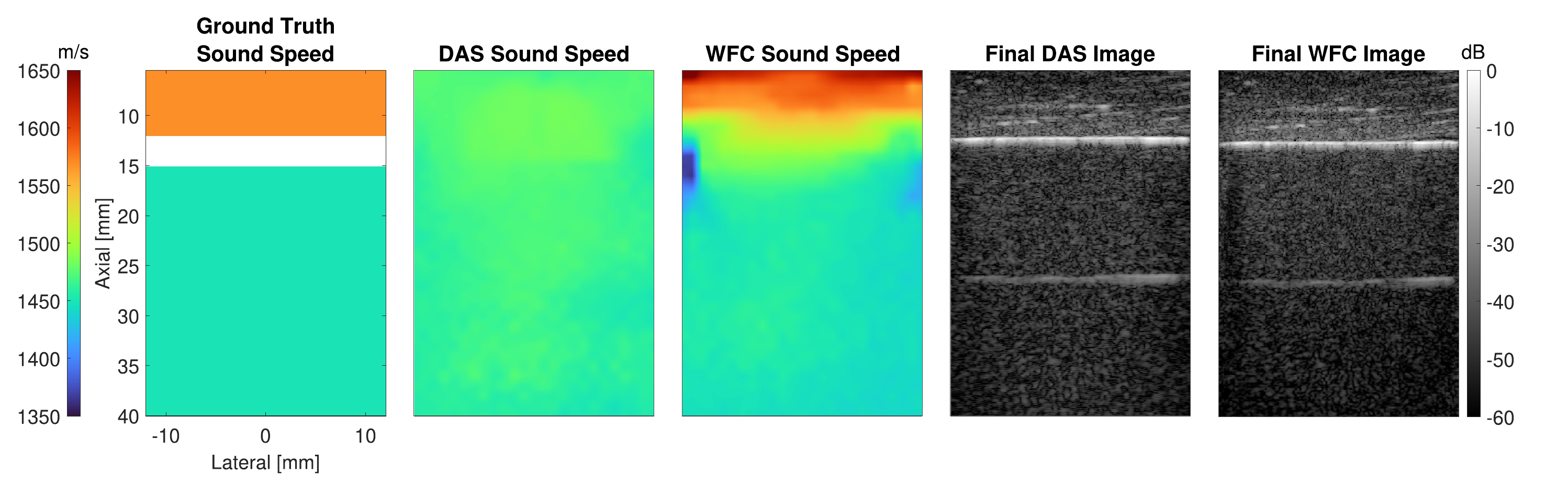}
    \caption{Sound speed estimations for an acquisition with chicken over a constant sound speed phantom. The ground truth sound speeds based on calibration measurements or phantom specifications are shown on the left. To avoid uncertainty in the boundary depth, a small region is excluded about the boundary between the chicken and phantom (shown in white). The chicken region appears clearly in the WFC sound speed estimation, with much lower bias compared to the DAS estimate, and the phantom region has lower error using WFC compared to DAS.}
    \label{fig:phantom_ats}
\end{figure*}

The sound speed estimates for the chicken overlying the ATS 549 phantom are shown in Figure \ref{fig:phantom_ats}, with errors listed in Table \ref{tab:error_phantom}. The measured sound speeds for the meat and phantom regions are visualized on the leftmost plot in Figure \ref{fig:phantom_ats}, with the depth around the boundary excluded (shown in white) due to the exact location of the boundary being unknown. Similarly to the abdominal wall simulations, the different sound speed regions are more visibly distinguishable in the WFC estimate compared to DAS, where the chicken region is more difficult to discern from the phantom region. Quantitatively, the bias is much higher in the chicken region for DAS than WFC. The standard deviation in the WFC sound speed estimation error, however, is higher than DAS due to the estimated sound speed decreasing in value towards the meat-phantom boundary. In the phantom region, which contains reverberation clutter from the chicken layer and phantom boundary, the DAS estimate has more than 10\,m/s of bias, similar to previous sound speed estimation literature \cite{Jakovljevic2018Local}. Conversely, the WFC estimate has lower bias in this region, although the standard deviation is slightly higher due to the drop in sound speed visible at the edges around 15 to 20\,mm in depth. 

\begin{table}[ht!]
    \centering
    \caption{Phantom Sound Speed Estimation Error}
    \label{tab:error_phantom}
    \begin{tabular}{c|c|c|c|c}
        \textbf{Medium} & \textbf{Beamformer} & \textbf{Bias} & \textbf{Std. Dev.} & \textbf{MAE} \\ 
        \hline
         \multirow{2}{*}{Chicken} & DAS & -92.74 & 4.84 & 92.74 \\
         & WFC & 8.90 & 42.10 & 33.74 \\
         \hline
         \multirow{2}{*}{ATS 549} & DAS & 14.24 & 6.00 & 14.25 \\
         & WFC & 0.76 & 11.78 & 7.25 \\
    \end{tabular}
\end{table}

\begin{figure*}[ht!]
    \centering
    \includegraphics[scale=0.37, trim={4mm, 3mm, 17mm, 5mm}, clip=true]{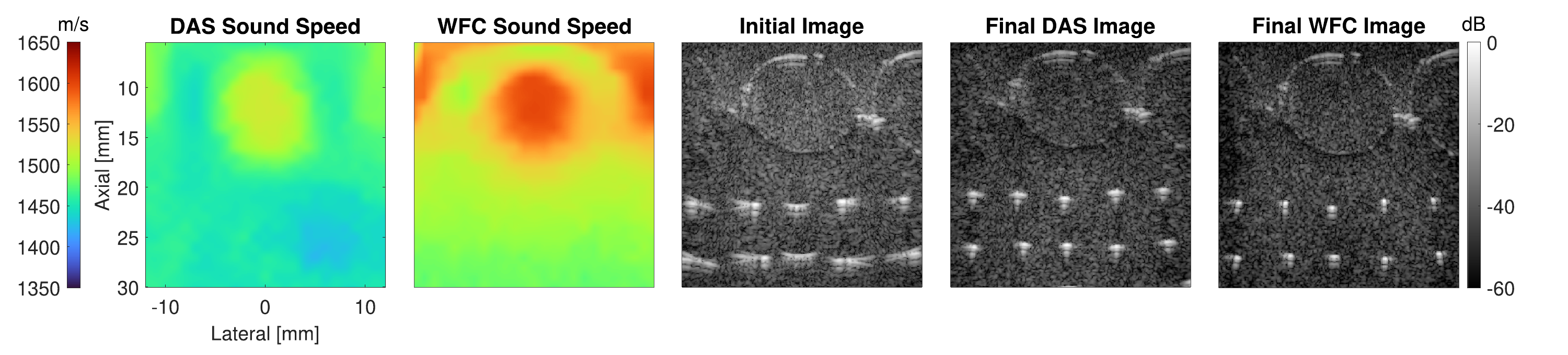}
    \caption{Sound speed estimation and aberration correction comparisons for a fabricated alcohol-gelatin phantom. The difference between the inclusion and background sound speeds is around 25\,m/s higher in the WFC model compared to the DAS model, and the background sound speed is more uniform for WFC compared to DAS. The wire target resolution improves with aberration correction compared to a constant sound speed, with further improvements when WFC is used instead of DAS.}
    \label{fig:phantom_impact}
\end{figure*}

The estimates for the fabricated alcohol-gelatin phantom are shown in Figure \ref{fig:phantom_impact}. Greater differentiation between the background and inclusion sound speeds are apparent in the WFC sound speed map, where the sound speed contrast is 25\,m/s higher than with DAS. The inclusion in the sound speed map also appears more circular in the WFC estimate. 
The eccentricity of the central lesion 
is 0.68 for DAS and 0.27 for WFC. The oblong inclusion shape for the DAS estimation is consistent with our previous experiments with this phantom and the straight-ray model \cite{Simson2025Ultrasound}.

Although the ground truth sound speed map is unknown for this phantom, the effect of the difference in sound speed estimation between the two models can be observed in the wire targets in the final beamformed images. 
While both models improve the B-mode images, indicating more accurate sound speed modeling, the wire targets in the image based on the WFC model have an average lateral FWHM of 0.25\,mm, while the DAS model has a lateral FWHM of 0.49\,mm. Furthermore, the wire target brightness is higher relative to the background brightness in the WFC model. The difference between the maximum brightness of the point targets and the surrounding background is approximately 40.9\,dB for the image beamformed with a constant sound speed, 45.2\,dB for the image using the DAS model, and 47.0\,dB for the image using the WFC model. Overall, the general improvement in image quality indicates that the WFC-estimated sound speed map has higher accuracy compared to the DAS-estimated sound speed map. 

\subsection{\textit{In Vivo}}
\begin{figure*}[ht!]
    \centering
    \begin{subfigure}[b]{0.99\textwidth}
        \centering
        \includegraphics[width=\linewidth, trim={0mm, 41mm, 0mm, 52mm}, clip=true]{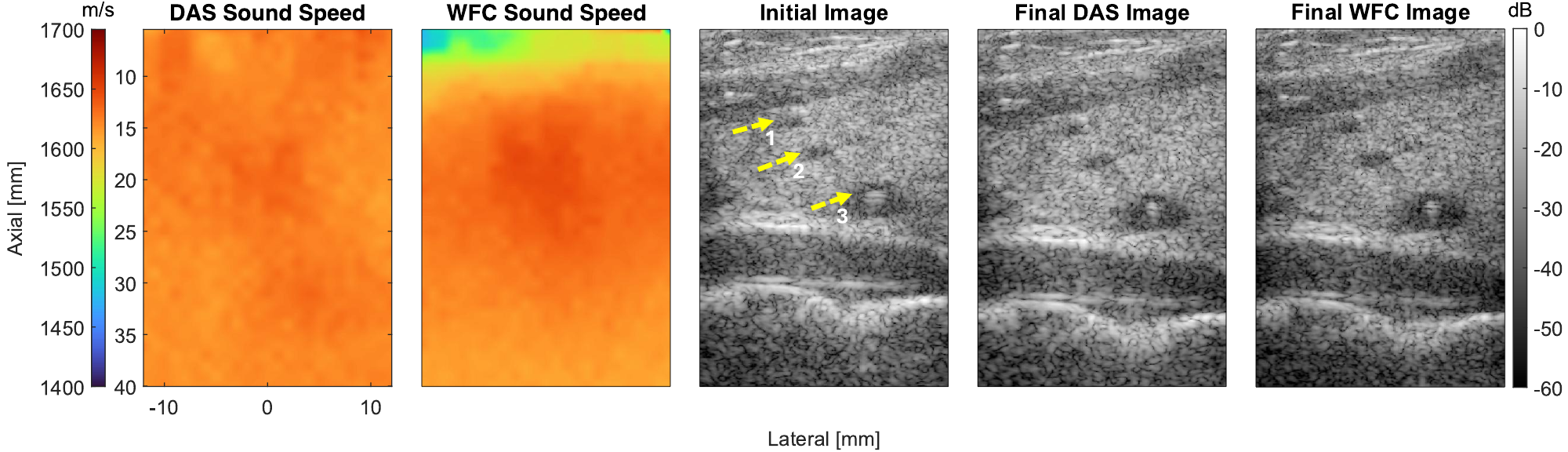}
        \caption{}
        \label{fig:invivo_1}
    \end{subfigure}
    \begin{subfigure}[b]{0.99\textwidth}
        \centering
        \includegraphics[width=\linewidth, trim={0mm, 44mm, 0mm, 50mm}, clip=true]{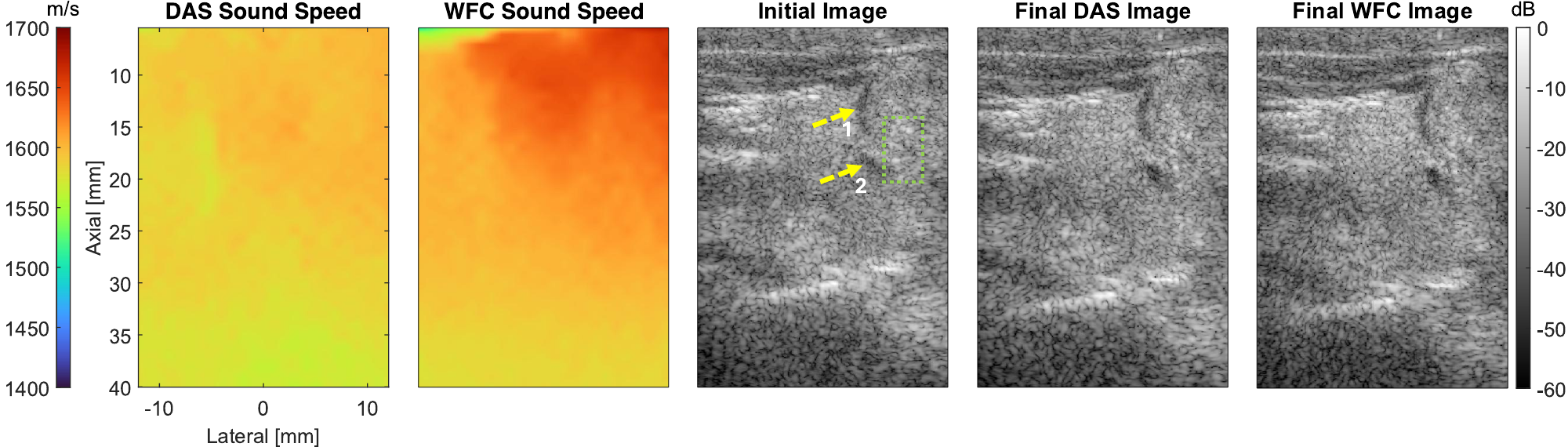}
        \caption{}
        \label{fig:invivo_3}
    \end{subfigure}
    \caption{Sound speed estimates and aberration corrected images for thyroid acquisitions from (a) Subject 1 and (b) Subject 2. Regions of interest for gCNR calculations are indicated by arrows. The contrast in hypoechoic and anechoic regions improves more greatly for the WFC model to compared to the straight-ray DAS model. The stronger scatterers exhibit improvements in resolution, especially when compared to beamforming with a constant sound speed, such as the scatterer in inclusion 3 in (a) or the scatterers forming the boundary of the structure in the green box in (b).}
    \label{fig:invivo}
\end{figure*}

Figure \ref{fig:invivo} depicts comparisons of representative sound speed estimates and beamformed images for thyroid acquisitions in two subjects at different scanning positions. Although the ground truth sound speeds are unknown, the images using local sound speed maps show resolution and contrast improvements from aberration correction compared to beamforming with an optimized constant sound speed (optimized based on the process in Section~\ref{sxn:implementation}). Contrast improvements are detailed in Table \ref{tab:invivo_gcnr}, with regions of interest indicated by arrows in Figure \ref{fig:invivo}. The gCNR comparisons show that the WFC model achieves greater inclusion separability in hypoechoic and anechoic regions. Additionally, the WFC model results in improved resolution in the final aberration-corrected images. In Figure \ref{fig:invivo_1}, the scatterer located in the center of the third inclusion has a lateral FWHM of 1.39\,mm when beamforming with an optimized constant sound speed, 0.72\,mm with DAS model, and 0.69\,mm with the WFC model. Similarly, in Figure \ref{fig:invivo_3}, the high amplitude scatterers located in the green box have a lateral FWHM of 0.38\,mm with the WFC model, whereas lateral FWHM is 0.4\,mm with the DAS model and 0.52\,mm with the optimized constant sound speed, respectively.

\begin{table}[hb!]
    \centering
    \caption{\textit{In Vivo} Thyroid gCNR Comparisons}
    \label{tab:invivo_gcnr}
    \begin{tabular}{c|c|c|c}
        \textbf{Region} & \textbf{Initial gCNR} & \textbf{DAS gCNR} & \textbf{WFC gCNR} \\ 
        \hline
        Fig. \ref{fig:invivo_1} Inclusion 1 & 0.68 & 0.66 & 0.84 \\
        Fig. \ref{fig:invivo_1} Inclusion 2 & 0.67 & 0.82 & 0.86 \\
        Fig. \ref{fig:invivo_1} Inclusion 3 & 0.84 & 0.92 & 0.93 \\
        \hline
        Fig. \ref{fig:invivo_3} Inclusion 1 & 0.81 & 0.87 & 0.86 \\
        Fig. \ref{fig:invivo_3} Inclusion 2 & 0.66 & 0.76 & 0.85 \\
    \end{tabular}
\end{table}

\begin{figure*}[ht!]
    \centering
    \includegraphics[width=\linewidth, trim={0mm, 64mm, 0mm, 65mm}, clip=true]{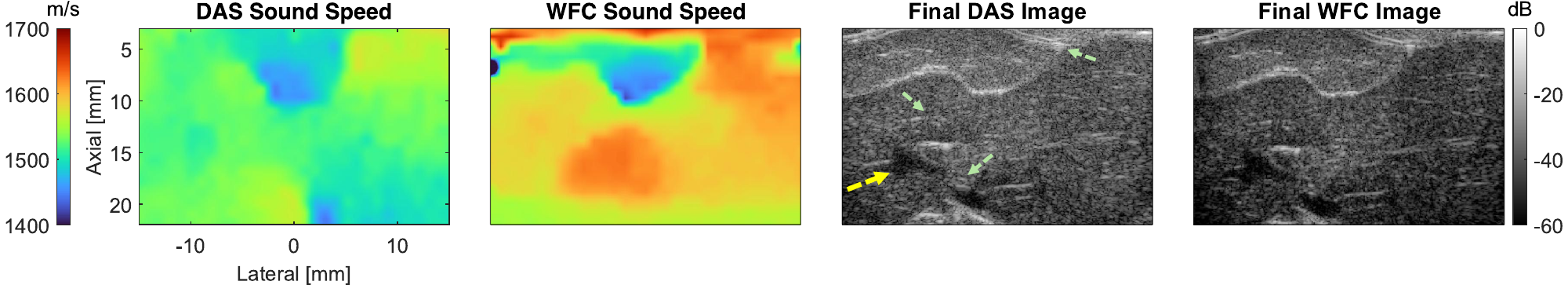}
    \caption{Sound speed estimation and aberration correction comparisons for an \textit{in vivo} rat liver acquisition. The transition between the abdominal wall and the liver region appears more clearly in the WFC sound speed estimate compared to the DAS estimate. Most of the tissue structure visible in the upper left of the beamformed images appears as a similarly shaped region with lower sound speed in the WFC estimate. In contrast, the DAS estimate presents moderately similar anatomy with a lower sound speed centered at approximately 0\,mm laterally. The WFC aberration corrected image shows improved resolution in locations indicated by green arrows and improved contrast in the region indicated by a yellow arrow.}
    \label{fig:invivo_rat}
\end{figure*}

Sound speed estimates for a rat liver acquisition with greater tissue heterogeneity are shown in Figure \ref{fig:invivo_rat}. The measured average ground truth sound speed in the liver was 1562\,m/s, resulting in an overall bias of -40\,m/s for the DAS sound speed estimate and 11\,m/s for the WFC sound speed estimate. The WFC estimate appears to distinguish different tissue regions better than the DAS estimate, with the abdominal wall above 5\,mm appearing distinct from the rest of the liver. Additionally, the liver lobe in the upper left of the B-mode image appears as a similarly-shaped region of lower sound speed in the WFC sound speed estimate, while only moderate anatomical similarity is obtained with the DAS estimate. The aberration-corrected WFC image demonstrates improved image quality compared to the DAS image as well, especially on the left side. The largest hypoechoic region indicated by a yellow arrow on the first B-mode image in Figure \ref{fig:invivo_rat} has a gCNR of 0.93 for the WFC image and a gCNR of 0.87 for the DAS image. 
The lateral FWHM for the stronger scatterers indicated by green arrows between -5 and -10\,mm laterally and 10 and 20\,mm axially decreases up to 0.18\,mm in the WFC image compared to the DAS image. On the upper right, the strong scatterer indicated by a green arrow at 6\,mm laterally and 4\,mm axially shows a reduction in lateral FWHM from 0.82\,mm with DAS to 0.61\,mm with WFC.


\section{Discussion}
\subsection{Sound Speed Estimation and Aberration Correction Enhancements}
The major improvements provided by the WFC model with respect to DAS beamforming with a straight-ray model include improved sound speed accuracy and reconstruction of complex tissue structures.
By simulating the wave propagation through a heterogeneous medium, refraction and diffraction of the waves are included in the wave propagation model, improving upon our prior DAS straight-ray model. In our prior model \cite{Simson2025Ultrasound}, we noted that the straight-ray model does not account for refraction or diffraction, resulting in errors in sound speed estimation. However, in the prior model yielded good aberration correction and image quality despite these errors due to the ill-posedness of the problem; that is, there are many solutions to the sound speed estimates that resulted in low CMPE. In our present WFC model, the use of improved modeling of wave propagation reduces the attribution of model error, instead of aberration, to the CMPE loss funciton. This is what leads to more accurate estimation of the speed of sound, especially when larger sound speed variations are present, as demonstrated by the shape of the inclusions in Figures \ref{fig:abd_inclusion} and \ref{fig:phantom_impact}. While the ground truth sound speeds were not availabe for this inclusion phantom, the material composition formulas used in Ali et al. \cite{Ali2023Sound} suggest the target sound speeds were 1650 and 1515\,m/s for the inclusions and background, respectively. Although these sound speeds cannot be guaranteed\footnote{Anecdotally, it has been observed that gelatin-alcohol phantoms of different composition will diffuse into each other, resulting in a change in sound speed.}, a higher sound speed estimate can be observed in both the inclusion and background of this phantom in the WFC model compared to DAS straight-ray model. 

While sound speed estimates from the straight-ray model can result in visible improvement to the final aberration-corrected images, a more accurate sound speed map can further improve image quality, as well as provide more reliable sound speed estimates for use in quantitative biomarkers of disease. Additionally, due to the spatiotemporal matched filtering from the correlation of an idealized forward-propagated transmit wavefield and the backward-propagated actual received wavefield, the effect of diffuse clutter can be reduced using WFC, which can improve the fidelity of the ultrasound signals that would otherwise be adversely affected by reverberation noise. This results in more accurate sound speed estimations and imporved image quality in cluttered scenarios, such as the reverberation shown in the lumen of the vasculature in Fig.~\ref{fig:invivo}a. However, this spatiotemporal matched filtering only occurs with the one-way propagation simulation method used here. A two-way propagation model, such as k-Wave or Fullwave, will match the reverberation signal from propagation of the reflected waves and is not expected to cancel the multiple reflections. 

There are a few anomalies worth noting with the sound speed estimates in the WFC model. First, the small artifacts at the corners of the estimates in the two simulation cases in Figure \ref{fig:abd_sims} and the sound speed dips on the edges of the meat-layer phantom in Figure \ref{fig:phantom_ats} do not consistently appear for all sources of data, but are also not uncommon. These artifacts can potentially be mitigated through different regularization parameters, changes to the field of view in the loss calculation or speed of sound grids, or different initialization parameters. Second, the worse sound speed estimates found in the bottom layer of Figure \ref{fig:abd_sims}(b) compared to the DAS model was an uncommon result. In the majority of simulation and experimental data containing heterogeneous distributions of speed of sound, we found the WFC model to more accurately estimate the speed of sound in terms of bias and MAE.

\subsection{Initialization and Pulse-Echo Depth-Velocity Ambiguity}
Due to the extreme non-convexity and ill-posedness of the optimization problem, initialization greatly affects the global (mean) sound speed and, consequently, the bias in the final sound speed map. The effect of the initialization on the final global sound speed offset has been demonstrated for other pulse-echo local sound speed estimators, such as Figure 10 in \cite{Jaeger2022PulseEcho} and Figure 5 in \cite{Ali2023Sound}. While the exact global sound speed---and thus the spatial locations of structures in the beamformed image---heavily depends on the constant sound speed used for initialization, aberration is still corrected to a similar degree in the final image using the estimated sound speed. Different initializations, coupled with differences in the field of view, can also affect the final sound speed estimation results, as seen when comparing the fabricated inclusion phantoms in Figure \ref{fig:phantom_impact} and the same phantom estimates in \cite{Simson2025Ultrasound} with a different field of view and initialization parameters. This phenomenon is caused by depth-velocity ambiguity in pulse-echo measurements, where the exact location of objects in a medium cannot be determined if the sound speed is unknown. This ambiguity is presented in theoretical detail in \cite{Podkowa2020Convolutional}, and similar occurrences of depth-velocity ambiguity in iterative sound speed estimation heavily influenced by initialization are discussed in \cite{Ali2023Sound} and \cite{Ali2023Distributed}. Although depth-velocity ambiguity can make the estimated sound speed values  less reliable for direct use in downstream applications, the relative change in sound speed in the map is still useful for applications like aberration correction. Regardless of the location of features due to the global sound speed offset, resolution and contrast still improve from accounting for the relative time delay changes caused by changing sound speed in media. 

An approach to reduce the effects of initialization is calibration (which can additionally account for modeling errors). Tomographic inversion approaches to estimating sound speed use phase shift calibration for different starting sound speeds to account for modeling errors \cite{Jaeger2022PulseEcho}. 
For an iterative sound speed estimation approach, \cite{Simson2025Ultrasound} instead utilizes an additional scalar time offset that can be calibrated for individual sources of data. This time offset essentially accounts for portions of the wave propagation process that are not modeled, such as the transducer lens and 3D effects. This time offset is chosen to minimize the sound speed bias for known homogeneous sound speed data for each data collection scenario, requiring separate calibration for both simulation and different scanned media to reduce estimation bias. While calibration approaches are more reliable in reducing the global bias, it either requires pre-calibration for each imaging scenario and starting sound speed, or knowing the sound speed in a medium, which could complicate standard clinical \textit{in vivo} imaging applications. In this work, both the initial sound speed and a time offset are swept prior to the initial beamforming iteration to find the pair of values that optimizes the CMPE. While this can result in lower estimation bias than simply sweeping for a starting sound speed, the initialization can have greater variation based on factors such as the exact medium imaged, field of view, or number of elements used during beamforming. 

However, to better address the global sound speed bias without calibration, future approaches would either need to overcome the depth-velocity ambiguity, or reliably initialize the sound speed and more accurately model the wave propagation in ultrasound scanner data. Better modeling of wave propagation through the transducer lens, as in \cite{Waasdorp2024Assessing}, could reduce the need for applying a time offset to account for propagation inaccuracies. Additionally, initialization could be posed as a separate problem to optimize by itself due to the strong influence of the initial sound speed on the final global sound speed of the estimate. For example, a hyperparameter tuning approach could be utilized to find the optimal parameters for initialization, as well as the rest of the optimization process. 

\subsection{Optimization Approach}
An advantage of auto-differentiation to optimize sound speed through gradient descent is the relative ease through which arbitrary loss criteria without an explicitly calculated gradient can be backpropagated to update a variable of interest, as long as all operations taken to calculate the loss with respect to that variable are backwards differentiable. Wavefield correlation as described in this work is a fully differentiable approach that accounts more accurately for wave propagation in a heterogeneous medium. Calculating Eikonal time delays is an alternative approach that accounts for refraction compared to the simpler straight ray approximation, but typical Eikonal equation solvers are not differentiable \cite{Sethian19993D}. At this time, we have found that the CMPE metric, which can be calculated in a differentiable way using separate sub-aperture split-step angular spectrum simulations, results in consistently better sound speed estimates compared to other image focusing metrics we have explored, such as those shown in \cite{Simson2025Ultrasound} and \cite{Frey2025UltraFlex}. 

The proposed optimization approach can have several disadvantages, some of which are similar to other sound speed estimators for pulse-echo ultrasound. Tracking all operations and necessary variable values to perform software auto-differentiation requires a considerable amount of memory, and can easily require high-end GPUs with more vRAM when utilizing GPU code acceleration. While DAS beamformers can benefit more from the GPU acceleration and have fewer operations to track, the angular spectrum simulation used for WFC must run sequentially in depth, resulting in longer runtimes, and more computational steps result in higher memory needs. However, the style of optimization at this stage of development must be performed offline anyway due to the gradient-descent-style iterative loss minimization. To reduce the runtime by requiring fewer iterative steps, a heterogeneous local sound speed map obtained from a real-time sound speed estimator could be used to initialize the optimization closer to the actual sound speed. Additionally, loss metrics that are faster to calculate than CMPE could be used for some or all parts of the iteration process, although this may come at the expense of performance. 

Similar to other sound speed estimators, a few components in the algorithm require manual tuning or adjustment. Regularization has a large impact on the morphology of the final result due to the ill-posedness of the pulse-echo local estimation problem. Simply optimizing for an image metric by itself often does not result in a physically plausible sound speed map, especially in cluttered data, due to the large rapid variations or extreme values that will develop in the sound speed map. Regularization, therefore, must be added to the optimized loss with the exact weight tuned for different media to enforce a balance in realistic smoothness and sound speed variation. Additionally, 
the stopping criterion is not standardized in this work and was set to a fixed number of iterations for which the sound speed estimation settled to stable values. Instead, an adaptive stopping criterion can be devised in the future based on the loss function, which could further reduce the runtime. 

\section{Conclusion}
We present a method for local sound speed imaging that improves estimation in challenging heterogeneous media compared to a state-of-the-art approach, especially in cases with reverberation clutter and large changes in sound speed. The proposed estimator optimizes sound speed by implementing the WFC beamformer---which involves simulating transmit and receive wavefield propagation---using auto-differentiation software. A total-variation regularized CMPE loss is calculated from beamformed data, and a local sound speed map is iteratively updated using gradient descent, with the backpropagated gradient automatically computed through software. We compare sound speed estimates and aberration correction for the proposed beamformer with a previously-published straight-ray DAS approach, showing larger estimated sound speed variation between different media regions, decreased estimation error, and improved image resolution and contrast. These results are promising for extending pulse-echo sound speed estimation to more challenging \textit{in vivo} imaging situations with significant clutter and sound speed changes. 

\section*{Acknowledgment}
The authors would like to thank Michael Jaeger, Di Xiao, Pat de la Torre, and Alfred Yu for their helpful discussions and insights regarding the behavior of speed of sound in custom made phantoms.

\bibliographystyle{IEEEtran}
\bibliography{IEEEabrv,abbrev,references}

@STRING{IEEE_J_UFFC       = "{IEEE} Trans. Ultrason., Ferroelectr., Freq. Control"}

@STRING{IEEE_J_MI         = "{IEEE} Trans. Med. Imag."}

@string{IEEE_J_CI="{IEEE} Trans. Comput. Imag."}

@string{IEEE_J_US="{IEEE} Trans. Ultrason."}

@string{TMI="IEEE Trans. Med. Imaging"}

@string{UMB="Ultrasound Med. Biol."}

@string{JASA="J. Acoust. Soc. Am."}

@string{UI="Ultrason. Imaging."}

@string{SPIE="Proc. SPIE"}

@string{PMB="Physics Med. Biol."}

@string{IUS_2018="Proc. 2018 {IEEE} Int. Ultrason. Symp. (IUS)"}

@string{IUS_2020="Proc. 2020 {IEEE} Int. Ultrason. Symp. (IUS)"}

@string{IUS_2023="Proc. 2023 {IEEE} Int. Ultrason. Symp. (IUS)"}

@string{IUS_2024="Proc. 2024 {IEEE} Int. Ultrason. Symp. (IUS)"}

@STRING{PMB="Phys. Med. Biol."}

@STRING{IJCARS="Int. J Comput. Assist. Radiol. Surg."}

@STRING{UI="Ultrason. Imaging"}

@STRING{MI_UIT_2023="Med. Imag. 2023: Ultrason. Imag. Tomogr."}

@string{NDTE="NDT E Int."}

@string{MIA="Med. Imag. Anal."}

@string{IOPP="IOP Publ."}

@STRING{GP="Geophysics"}

@ARTICLE{Ali2023Aberration,
	author = "R.~Ali and T.~Brevett and L.~Zhuang and H.~Bendjador and A.~S.~Podkowa and S.~Hsieh and W.~Simson and S.~J.~Sanabria and C.~Herickhoff and J.~J.~Dahl",
	title = "Aberration Correction in Diagnostic Ultrasound: A Review of the Prior Field and Current Directions",
	journal = "Z. Med. Phys.",
	year = 2023,
	volume = 33,
	number = 3,
	pages = "267--291"}

@article{Deng2017Quantifying,
  title={Quantifying image quality improvement using elevated acoustic output in {B}-mode harmonic imaging},
  author={Deng, Yufeng and Palmeri, Mark L and Rouze, Ned C and Trahey, Gregg E and Haystead, Clare M and Nightingale, Kathryn R},
  journal=UMB, 
  volume={43},
  number={10},
  pages={2416--2425},
  year={2017},
  publisher={Elsevier}
}

@article{Smereczynski2018Pitfalls, 
    title={Pitfalls in ultrasound imaging of the stomach and the intestines}, 
    author={Smereczy\'{n}ski, Andrzej and Kołaczyk, Katarzyna}, 
    journal="J. Ultrason.", 
    volume={18}, 
    number={74}, 
    pages={207--211}, 
    year={2018}
}

@ARTICLE{Jaeger2015Full,
  author={Jaeger, M. and Robinson, E. and Akar\c{c}ay, H. G. and Frenz, M.},
  title={Full correction for spatially distributed speed-of-sound in echo ultrasound based on measuring aberration delays via transmit beam steering},
  journal=PMB,
  year={2015},
  volume={60},
  number={11},
  pages={4497-4515}
}

@article{Jaeger2015Computed,
    title = {Computed Ultrasound Tomography in Echo Mode for Imaging Speed of Sound Using Pulse-Echo Sonography: Proof of Principle},
    journal = UMB,
    volume = {41},
    number = {1},
    pages = {235-250},
    year = {2015},
    issn = {0301-5629},
    doi = {https://doi.org/10.1016/j.ultrasmedbio.2014.05.019},
    urlj = {https://www.sciencedirect.com/science/article/pii/S0301562914003500},
    author = {Michael Jaeger and Gerrit Held and Sara Peeters and Stefan Preisser and Michael Grünig and Martin Frenz},
}

@article{Stahli2020Improved,
    title = {Improved forward model for quantitative pulse-echo speed-of-sound imaging},
    journal = {Ultrasonics},
    volume = {108},
    pages = {106168},
    year = {2020},
    author = "Patrick {St\"ahli} and Maju Kuriakose and Martin Frenz and Michael Jaeger"}

@article{Rau2021Speed,
    author = {Rau, Richard and Schweizer, Dieter and Vishnevskiy, Valery and Goksel, Orcun},    
    title = {Speed-of-sound imaging using diverging waves},
    journal = IJCARS,
    volume = {16}, 
    number = {7},    
    pages = {1201-1211},
    year = {2021},
    urlj = {https://doi.org/10.1007/s11548-021-02426-w},
    doi = {10.1007/s11548-021-02426-w},
}

@article{Sanabria2018Spatial,
    doi = {10.1088/1361-6560/aae2fb},
    urlj = {https://doi.org/10.1088/1361-6560/aae2fb},
    year = {2018},
    month = {Oct.},
    publisher = IOPP,
    volume = {63},
    number = {21},
    pages = {215015},
    author = {Sanabria, Sergio J and Ozkan, Ece and Rominger, Marga and Goksel, Orcun},
    title = {Spatial domain reconstruction for imaging speed-of-sound with pulse-echo ultrasound: simulation and in vivo study},
    journal = PMB,
}

@article{Jakovljevic2018Local,
    author = {Jakovljevic, Marko and Hsieh, Scott and Ali, Rehman and Chau Loo Kung, Gustavo and Hyun, Dongwoon and Dahl, Jeremy J.},
    title = "{Local speed of sound estimation in tissue using pulse-echo ultrasound: Model-based approach}",
    journal = JASA,
    volume = {144},
    number = {1},
    pages = {254-266},
    year = {2018},
    month = {July}}

@ARTICLE{Ali2022Local,
  author={Ali, Rehman and Telichko, Arsenii V. and Wang, Huaijun and Sukumar, Uday K. and Vilches-Moure, Jose G. and Paulmurugan, Ramasamy and Dahl, Jeremy J.},
  journal=IEEE_J_UFFC,
  title={Local sound speed estimation for pulse-echo ultrasound in layered media}, 
  year={2022},
  volume={69},
  number={2},
  pages={500-511},
  doi={10.1109/TUFFC.2021.3124479}
}

@ARTICLE{Ali2022Distributed,
  author={Ali, Rehman and Brevett, Thurston and Hyun, Dongwoon and Brickson, Leandra L. and Dahl, Jeremy J.},
  journal=IEEE_J_UFFC, 
  title={Distributed Aberration Correction Techniques Based on Tomographic Sound Speed Estimates}, 
  year={2022},
  volume={69},
  number={5},
  pages={1714-1726},
  doi={10.1109/TUFFC.2022.3162836}}

@INPROCEEDINGS{Ali2018Distributed,
  author={Ali, Rehman and Dahl, Jeremy J.},
  booktitle=IUS_2018, 
  title={Distributed phase aberration correction techniques based on local sound speed estimates}, 
  year={2018},
  volume={},
  number={},
  pages={1-4},
  doi={10.1109/ULTSYM.2018.8580139}
}

@INPROCEEDINGS{Beuret2020Refraction,
  author={Beuret, Samuel and Perdios, Dimitris and Thiran, Jean-Philippe},
  booktitle=IUS_2020, 
  title={Refraction-Aware Integral Operator for Speed-of-Sound Pulse-Echo Imaging}, 
  year={2020},
  volume={},
  number={},
  pages={1-4},
  doi={10.1109/IUS46767.2020.9251601}}

@article{Masoy2005Iteration,
    author = {Måsøy, Svein-Erik and Varslot, Trond and Angelsen, Bjørn},
    title = {Iteration of transmit-beam aberration correction in medical ultrasound imaging},
    journal = JASA,
    volume = {117},
    number = {1},
    pages = {450-461},
    year = {2005},
    month = {01},
    issn = {0001-4966},
    doi = {10.1121/1.1823213},
    urlj = {https://doi.org/10.1121/1.1823213},
    eprint = {https://pubs.aip.org/asa/jasa/article-pdf/117/1/450/8094927/450_1_online.pdf},
}

@ARTICLE{Krishnan1996Adaptive,
  author={Krishnan, S. and Pai-Chi Li and O'Donnell, M.},
  journal=IEEE_J_UFFC, 
  title={Adaptive compensation of phase and magnitude aberrations}, 
  year={1996},
  volume={43},
  number={1},
  pages={44-55},
  doi={10.1109/58.484462}}

@article{Wang2025Pixel,
    title = {Pixel-responsive optimization beamforming method for ultrasound transcranial imaging},
    journal = MIA,
    volume = {106},
    pages = {103762},
    year = {2025},
    issn = {1361-8415},
    doi = {https://doi.org/10.1016/j.media.2025.103762},
    urlj = {https://www.sciencedirect.com/science/article/pii/S1361841525003081},
    author = {Junyi Wang and Tianhua Zhou and Gaobo Zhang and Boyi Li and Xin Liu and Dean Ta},
}

@INPROCEEDINGS{Ali2023Iterative,
  author={Ali, Rehman and Mitcham, Trevor and Singh, Melanie and Bouchard, Richard and Doyley, Marvin and Dahl, Jeremy and Duric, Nebojsa},
  booktitle=IUS_2023, 
  title={Iterative Sound Speed Tomography for Distributed Aberration Correction}, 
  year={2023},
  volume={},
  number={},
  pages={1-4},
  doi={10.1109/IUS51837.2023.10306695}}

@Article{Zengqiu2024Iterative,
    AUTHOR = {Zengqiu, Yuchen and Wu, Wentao and Xiao, Ling and Zhou, Erlei and Cao, Zheng and Hua, Jiadong and Wang, Yue},
    TITLE = {Iterative Pulse–Echo Tomography for Ultrasonic Image Correction},
    JOURNAL = {Sensors},
    VOLUME = {24},
    YEAR = {2024},
    NUMBER = {6},
    ARTICLE-NUMBER = {1895},
    URLj = {https://www.mdpi.com/1424-8220/24/6/1895},
    PubMedID = {38544158},
    ISSN = {1424-8220},
    DOI = {10.3390/s24061895}
}

@ARTICLE{Ali2023Sound,
  author={Ali, Rehman and Mitcham, Trevor M. and Singh, Melanie and Doyley, Marvin M. and Bouchard, Richard R. and Dahl, Jeremy J. and Duric, Nebojsa},
  journal=IEEE_J_CI, 
  title={Sound Speed Estimation for Distributed Aberration Correction in Laterally Varying Media}, 
  year={2023},
  volume={9},
  number={},
  pages={367-382},
  doi={10.1109/TCI.2023.3261507}}

@ARTICLE{Simson2025Ultrasound,
	author = "W.~A.~Simson and L.~Zhuang and B.~N.~Frey and S.~J.~Sanabria and J.~J.~Dahl and D.~Hyun",
	title = "Ultrasound Autofocusing: Common Midpoint Phase Error Optimization via Differentiable Beamforming",
	journal = TMI,
	year = 2026,
	volume = 45,
	number = 2,
	pages = "681--692"}

@INPROCEEDINGS{Zhuang2024Simultaneous,
  author={Zhuang, Louise and Brevett, Thurston and Hyun, Dongwoon and Dahl, Jeremy},
  booktitle=IUS_2024, 
  title={Simultaneous Reverberation Noise Reduction and Aberration Correction Using Wavefield Correlation}, 
  year={2024},
  volume={},
  number={},
  pages={1-5},
  doi={10.1109/UFFC-JS60046.2024.10793721}}

@INPROCEEDINGS{Schwab2018Full,
  author={Schwab, Hans-Martin and Schmitz, Georg},
  booktitle=IUS_2018, 
  title={Full-Wave Ultrasound Reconstruction with Linear Arrays Based on a Fourier Split-Step Approach}, 
  year={2018},
  volume={},
  number={},
  pages={1-4},
  doi={10.1109/ULTSYM.2018.8580147}}

@article{Schliecher2008shot,
    author = "J.~Schleicher and J.~C.~Costa and A.~Novais",
    title = "A comparison of imaging conditions for wave-equation shot-profile migration",
    journal = GP,
    volume = 73,
    number = 6,
    pages = "s219-s227",
    year = 2008}

@article{Wang2013Transcranial,
    author = "T.~Wang and Y.~Jing",
    title = "Transcranial ultrasound imaging with speed of sound-based phase correction: a numerical study",
    journal = PMB,
    volume = 58,
    pages = "6663-–6681",
    year = 2013}

@article{Ali2021Fourier,
    author = {Rehman Ali},
    title ={Fourier-based Synthetic-aperture Imaging for Arbitrary Transmissions by Cross-correlation of Transmitted and Received Wave-fields},
    journal = UI,
    volume = {43},
    number = {5},
    pages = {282-294},
    year = {2021}}

@article{Rao2019Ultrasonic,
    author = "J.~Rao and A.~Saini and J.~Yang and M.~Ratassepp and Z.~Fan",
    title = {Ultrasonic imaging of irregularly shaped notches based on elastic reverse time migration},
    journal = NDTE,
    volume = {107},
    pages = {102135},
    year = {2019}}

@article{Sethian19993D,
    author = {James A. Sethian and A. Mihai Popovici},
    title = {3-{D} traveltime computation using the fast marching method},
    journal = GP,
    volume = {64},
    number = {2},
    pages = {516-523},
    year = {1999}}

@article{Jones2016Comparison,
    doi = {10.1088/0031-9155/61/1/23},
    urlj = {https://doi.org/10.1088/0031-9155/61/1/23},
    year = {2015},
    month = {Nov.},
    publisher = IOPP,
    volume = {61},
    number = {1},
    pages = {23},
    author = {Jones, Ryan M and Hynynen, Kullervo},
    title = {Comparison of analytical and numerical approaches for {CT}-based aberration correction in transcranial passive acoustic imaging},
    journal = PMB,
}

@article{Claerbout1971Toward,
    author = {Claerbout, Jon F.},
    title = "{Toward a unified theory of reflector mapping}",
    journal = GP,
    volume = {36},
    number = {3},
    pages = {467-481},
    year = {1971},
    month = {06},
    issn = {0016-8033},
    doi = {10.1190/1.1440185},
    urlj = {https://doi.org/10.1190/1.1440185},
    eprint = {https://pubs.geoscienceworld.org/seg/geophysics/article-pdf/36/3/467/3155298/467.pdf},
}

@article{Mallart1994Adaptive,
    author = {Mallart, Raoul and Fink, Mathias},
    title = {Adaptive focusing in scattering media through sound‐speed inhomogeneities: The van {Cittert} {Zernike} approach and focusing criterion},
    journal = JASA,
    volume = {96},
    number = {6},
    pages = {3721-3732},
    year = {1994},
    month = {12},
    issn = {0001-4966},
    doi = {10.1121/1.410562},
    urlj = {https://doi.org/10.1121/1.410562},
    eprint = {https://pubs.aip.org/asa/jasa/article-pdf/96/6/3721/11607792/3721_1_online.pdf},
}

@article{Mallart1991van,
    author = {Mallart, Raoul and Fink, Mathias},
    title = {The van {C}ittert–{Z}ernike theorem in pulse echo measurements},
    journal = {The Journal of the Acoustical Society of America},
    volume = {90},
    number = {5},
    pages = {2718-2727},
    year = {1991},
    month = {11},
    doi = {10.1121/1.401867},
    url = {https://doi.org/10.1121/1.401867},
    eprint = {https://pubs.aip.org/asa/jasa/article-pdf/90/5/2718/11448272/2718_1_online.pdf},
}

@ARTICLE{Pinton2009Heterogeneous,
  author={Pinton, Gianmarco F. and Dahl, Jeremy and Rosenzweig, Stephen and Trahey, Gregg E.},
  journal=IEEE_J_UFFC, 
  title={A heterogeneous nonlinear attenuating full-wave model of ultrasound}, 
  year={2009},
  volume={56},
  number={3},
  pages={474-488},
  doi={10.1109/TUFFC.2009.1066}
}

@article{Pinton2021Fullwave,
  title={A fullwave model of the nonlinear wave equation with multiple relaxations and relaxing perfectly matched layers for high-order numerical finite-difference solutions},
  author={Pinton, Gianmarco},
  journal={arXiv preprint arXiv:2106.11476},
  eprint={2106.11476},
  archivePrefix={arXiv},
  primaryClass={physics.med-ph},
  year={2021}
}

@ARTICLE{Brickson2021Reverberation,
  author={Brickson, Leandra L. and Hyun, Dongwoon and Jakovljevic, Marko and Dahl, Jeremy J.},
  journal=IEEE_J_MI, 
  title={Reverberation Noise Suppression in Ultrasound Channel Signals Using a {3D} Fully Convolutional Neural Network}, 
  year={2021},
  volume={40},
  number={4},
  pages={1184-1195},
  doi={10.1109/TMI.2021.3049307}}

@ARTICLE{RodriguezMolares2020Generalized,
  author={Rodriguez-Molares, Alfonso and Rindal, Ole Marius Hoel and D’hooge, Jan and Måsøy, Svein-Erik and Austeng, Andreas and Lediju Bell, Muyinatu A. and Torp, Hans},
  journal=IEEE_J_UFFC, 
  title={The Generalized Contrast-to-Noise Ratio: A Formal Definition for Lesion Detectability}, 
  year={2020},
  volume={67},
  number={4},
  pages={745-759},
  doi={10.1109/TUFFC.2019.2956855}}

@article{Telichko2022Noninvasive,
doi = {10.1088/1361-6560/ac4562},
urlj = {https://doi.org/10.1088/1361-6560/ac4562},
year = {2022},
month = {jan},
publisher = IOPP,
volume = {67},
number = {1},
pages = {015007},
author = {Telichko, Arsenii V and Ali, Rehman and Brevett, Thurston and Wang, Huaijun and Vilches-Moure, Jose G and Kumar, Sukumar U and Paulmurugan, Ramasamy and Dahl, Jeremy J},
title = {Noninvasive estimation of local speed of sound by pulse-echo ultrasound in a rat model of nonalcoholic fatty liver},
journal = PMB,
}

@ARTICLE{Zhuang2025Labeled,
  author={Zhuang, Louise and Ostras, Oleksii and Sode, Masashi and Simson, Walter and Hyun, Dongwoon and Santibanez, Francisco and Dahl, Jeremy and Pinton, Gianmarco},
  journal=IEEE_J_US, 
  title={Labeled numerical phantom of abdominal wall for wave-physics based ultrasound imaging: applications to image reconstruction}, 
  year={2025},
  volume={},
  number={},
  pages={1-1},
  doi={10.1109/TUSON.2025.3638314}
}

@article{Podkowa2020Convolutional,
    doi = {10.1088/1361-6560/ab6071},
    urlj = {https://doi.org/10.1088/1361-6560/ab6071},
    year = {2020},
    month = {Jan.},
    publisher = IOPP,
    volume = {65},
    number = {2},
    pages = {025003},
    author = {Podkowa, Anthony S and Oelze, Michael L},
    title = {The convolutional interpretation of registration-based plane wave steered pulse-echo local sound speed estimators},
    journal = PMB,
}

@inproceedings{Ali2023Distributed,
author = {Rehman Ali and Trevor Mitcham and Melanie Singh and Richard Bouchard and Jeremy Dahl and Marvin Doyley and Nebojsa Duric},
title = {{Distributed aberration correction in handheld ultrasound based on tomographic estimates of the speed of sound}},
volume = {12470},
booktitle = MI_UIT_2023,
editor = {Christian Boehm and Nick Bottenus},
organization = {International Society for Optics and Photonics},
publisher = {SPIE},
pages = {1247009},
year = {2023},
doi = {10.1117/12.2653935},
URLj = {https://doi.org/10.1117/12.2653935}
}

@article{Jaeger2022PulseEcho,
doi = {10.1088/1361-6560/ac96c6},
urlj = {https://doi.org/10.1088/1361-6560/ac96c6},
year = {2022},
month = {Oct.},
publisher = IOPP,
volume = {67},
number = {21},
pages = {215016},
author = {Jaeger, Michael and Stähli, Patrick and Korta Martiartu, Naiara and Salemi Yolgunlu, Parisa and Frappart, Thomas and Fraschini, Christophe and Frenz, Martin},
title = {Pulse-echo speed-of-sound imaging using convex probes},
journal = PMB,
}

@article{Frey2025UltraFlex,
    author = "B.~N.~Frey and D.~Hyun and W.~Simson and L.~Zhuang and H.~S.~Hashemi and M.~Schneider and J.~J.~Dahl",
    title = "{UltraFlex}: Iterative Model-Based Ultrasonic Flexible-Array Shape Calibration",
    journal = IEEE_J_UFFC,
    year = 2025,
    volume = 72,
    number = 11,
    pages = "1462--1475"}

@ARTICLE{Waasdorp2024Assessing,
  author={Waasdorp, Rick and Maresca, David and Renaud, Guillaume},
  journal=IEEE_J_UFFC, 
  title={Assessing Transducer Parameters for Accurate Medium Sound Speed Estimation and Image Reconstruction}, 
  year={2024},
  volume={71},
  number={10},
  pages={1233-1243},
  doi={10.1109/TUFFC.2024.3445131}}

@article{Kingma2017Adam,
      title={Adam: A Method for Stochastic Optimization}, 
      author={Diederik P. Kingma and Jimmy Ba},
      year={2017},
      journal={arXiv preprint arXiv:1412.6980},
      eprint={1412.6980},
      archivePrefix={arXiv},
      primaryClass={cs.LG},
      urlj={https://arxiv.org/abs/1412.6980}, 
}

@ARTICLE{Dahl2014Reverb,
  author = "J.~J.~Dahl and N.~M.~Sheth",
  title = "Reverberation Clutter from Subcutaneous Tissue Layers: Simulation and In Vivo Demonstrations",
  journal = UMB,
  year = 2014,
  volume = 40, 
  number = 4,
  pages = "714--726"}

\end{document}